\documentclass[paper]{JHEP3}
\usepackage{amssymb,enumerate}
\usepackage{amsmath}
\usepackage{bbm}
\usepackage{url}
\usepackage{cite}
\usepackage{graphics}
\usepackage{xspace}
\usepackage{epsfig}
\usepackage{subfigure}

\newcommand\POWHEG{{\tt POWHEG}}
\newcommand\POWHEGBOX{{\tt POWHEG BOX}}
\newcommand\PYTHIA{{\tt PYTHIA}}
\newcommand\POWHEGpPYTHIA{{\tt POWHEG+PYTHIA}}
\newcommand\HERWIG{{\tt HERWIG}}

\def\({\left(} 
\def\){\right)} 

\def\beq{\begin{equation}}
\def\beqn{\begin{eqnarray}}
\def\eeq{\end{equation}}
\def\eeqn{\end{eqnarray}}

\title{$W^+W^+$ plus dijet production in the \POWHEGBOX{}}
\vfill
\author{Tom Melia \\
Rudolf Peierls Centre for Theoretical Physics, 1 Keble Road, University of Oxford, UK\\
E-mail: \email{t.meliai1@physics.ox.ac.uk}}

\author{Paolo Nason \\
INFN, Sezione di Milano Bicocca, Italy\\
E-mail: \email{Paolo.Nason@mib.infn.it}}

\author{Raoul Rontsch \\
Rudolf Peierls Centre for Theoretical Physics, 1 Keble Road, University of Oxford, UK\\
E-mail: \email{r.rontsch1@physics.ox.ac.uk}}

\author{Giulia Zanderighi \\
Rudolf Peierls Centre for Theoretical Physics, 1 Keble Road, University of Oxford, UK\\
E-mail: \email{g.zanderighi1@physics.ox.ac.uk}}


\abstract{We present an implementation of the calculation of the
  production of $W^+W^+$ plus two jets at hadron colliders, at
  next-to-leading order (NLO) in QCD, in the \POWHEG{} framework, which is a
  method that allows the interfacing of NLO
  calculations to shower Monte Carlo programs.  This is the first
  $2\to 4$ process to be described to NLO accuracy within a shower
  Monte Carlo framework.  The implementation was built within the
  \POWHEGBOX{} package. We discuss a few technical improvements that
  were needed in the \POWHEGBOX{} to deal with the computer intensive
  nature of the NLO calculation, and argue that further improvements are
  possible, so that the method can match the complexity that is 
  reached today in NLO calculations.  We have
  interfaced our \POWHEG{} implementation with \PYTHIA{} and
  \HERWIG{}, and present some phenomenological results, discussing
  similarities and differences between the pure NLO and the
  \POWHEGpPYTHIA{} calculation both for inclusive and more exclusive
  distributions. We have made the relevant code available at the
  \POWHEGBOX{} web site.  \preprint{OUTP-11-34P}
}

\begin{document}

\section{Introduction}

With the increase in energy and luminosity of the LHC, accurate
predictions for high-multiplicity processes become necessary.  A lot
of effort has been devoted in recent years towards the calculation of
next-to-leading (NLO) corrections to various $2\to3$ and $2\to4$
scattering processes\footnote{As usual, in this counting we do not
  include the decay of heavy particles.} ~\cite{Bredenstein:2009aj,
  Bredenstein:2010rs, Bevilacqua:2009zn,
  Berger:2009zg,Berger:2009ep,Ellis:2009zw,KeithEllis:2009bu,
  Melnikov:2009wh,bbbb,tt2j,Berger:2010vm,Melia:2010bm,Denner:2010jp}.
Very recently, even dominant corrections to a $2\to 5$ process have
been computed~\cite{Berger:2010zx}.  When NLO predictions are
available, theoretical uncertainties are reduced compared to Born
level predictions, and more accurate comparisons with experimental
data become possible. However, NLO predictions describe the effect due
to at most one additional parton in the final state. This is quite far
from realistic LHC events, which involve a large number of particles
in the final state. For infrared safe, sufficiently inclusive
observables, NLO calculations provide accurate predictions, but this
is not the case for more exclusive observables that are sensitive to
the complex structure of LHC events.

A complementary approach is provided by parton shower event
generators, that generate realistic hadron-level events, but only with
leading logarithmic accuracy. In recent years, methods that include
the benefits of a NLO calculation together with a parton shower model
(an NLO+PS generator, from now on) have become available. Using these
methods one can thus generate exclusive, realistic events, maintaining
NLO accuracy for inclusive observables.  Two NLO+PS frameworks are
being currently used in hadron collider physics: {\tt
  MC@NLO}~\cite{Frixione:2002ik} and
\POWHEG~\cite{Nason:2004rx,Frixione:2007vw}.  In the past few years a
number of processes have been implemented in both frameworks. However,
most processes included so far are relatively simple $2\to 1$ or $2\to
2$ scattering processes, for which the one-loop correction can be
expressed in a closed, relatively simple analytic form.\footnote{Two
  noticeable exceptions are the \POWHEG{} implementations of two $2\to
  3$ processes: vector boson fusion Higgs
  production~\cite{Nason:2009ai}, and top pair production in
  association with one jet~\cite{Kardos}.}

No $2\to4$ process has been implemented so far in any NLO+PS
framework.  Tools to tackle processes of arbitrary complexity do
however exist.  A general computer framework for the \POWHEG{}
implementation of arbitrary NLO processes has been presented in
ref.~\cite{Alioli:2010xd}, the so called \POWHEGBOX{}. Within this
framework, one needs only to provide few ingredients: the phase-space
and flavour information, the (spin- and colour-correlated) Born,
real and virtual matrix elements for a given NLO process in order to
build a \POWHEG{} implementation of it.

Recent NLO calculations of processes of high multiplicity use
numerical methods to perform the reduction of tensor integrals 
``on the fly'' or to compute coefficients of master integrals in terms of
products of tree-level amplitudes. These methods allow the computation
of the virtual corrections to very complex processes. On the other
hand, these calculations become quite computer intensive.  The
computation of real radiation corrections also requires a considerable
CPU time, since one needs to integrate over a phase space of high
dimension. For instance, if $n$ on shell particles are produced at Born
level, the real radiation term involves an integration over $3n+1$
variables.

In this paper we present a \POWHEGBOX{} implementation of the QCD
production of $W^+W^+$+2 jets, including the leptonic decay of the $W$
bosons with spin correlations.  This is the first time that a $2\to 4$
process has been implemented in an NLO+PS framework.  NLO QCD
corrections to $W^+W^+$ production have been computed recently using
$D$-dimensional unitarity~\cite{Melia:2010bm}.  The production of
a $W^+W^+$ pair in hadronic collisions requires the presence of two
jets in the final state. Thus, in spite
of the presence of the these jets, there are no collinear or
soft divergences at the Born level.
As such, the process presents no complications due
to the need of a generation cut~\cite{Alioli:2010qp}.  However, since the NLO calculation
is computer intensive, a number of technical issues arise in \POWHEG{}
that are not present for simpler processes. We discuss some of these
issues in the present work, and find acceptable solutions for some of
them. We also show that further efficiency improvements are possible,
thus paving the road for the matching of NLO calculations and parton
showers for yet more complex processes.

\FIGURE[t]{
\includegraphics[angle=0,scale=0.6]{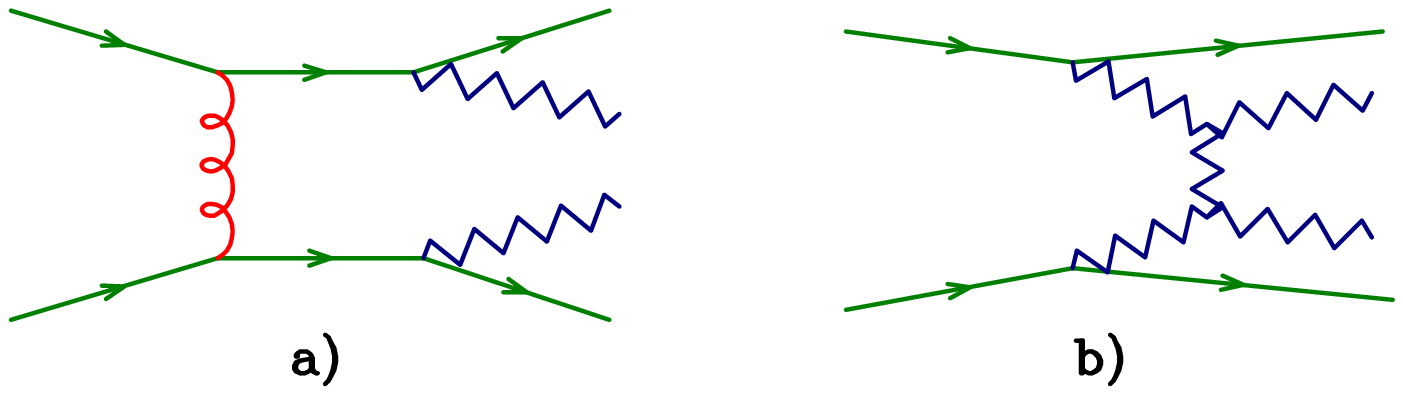}
\caption{Sample diagrams for a) QCD and b) electroweak production mechanisms of $W^+
  W^+$ plus dijet production.}
\label{fig:QCDandVBF}
} 

We consider in this work only the QCD production mechanism of the
$W^+W^+$+2 jets final state, i.e. the process involving one gluon
exchange and the direct emission of the $W^+$ pair from quark lines
(see fig.~\ref{fig:QCDandVBF}a).
A different production mechanism is given by the electroweak (EW) 
scattering process, where a colourless vector boson is exchanged in the
$t$-channel (see fig.~\ref{fig:QCDandVBF}b).
This second mechanism, although of higher order in the electroweak
coupling constant, is only
moderately smaller than the QCD one. We leave the \POWHEG{}
implementation of the EW production process, for which NLO QCD
corrections are also known~\cite{jager}, to a future publication.

The $W^+W^+$+2 jets process has a considerable phenomenological
interest.  It's LHC cross section, including the branching ratios to
electrons and muons, is around $6\;$fb at $7\;$TeV and $20\;$fb at $14\;$TeV.
It has a distinct signature of two same-sign leptons, missing energy
and two jets.  It therefore constitutes an important background to
studies of double-parton scattering~\cite{dps}, as well as to new
physics signatures that involve two same-sign leptons, such as
R-parity violating SUSY models~\cite{dreiner}, diquark production with
decay of the diquark to a pair of top quarks~\cite{han} or double
charged Higgs production~\cite{maalampi}. This work will make it
possible to have a more reliable generator of this SM background in
those physics studies, that currently use only LO (Leading Order)
shower Monte Carlo programs.

The remainder of this paper is organized as follows. 
In section~\ref{sec:WW} we discuss the process under study. 
In section~\ref{sec:tech} we present few technical issues related to
the implementation of the process in the
\POWHEGBOX{}~\cite{Alioli:2010xd} (more details are given in
appendix~\ref{Technicalities}).
In section~\ref{sec:res} we present physical results for some
kinematic distributions. We pay particular attention to where NLO
results differ mostly from the \POWHEG{} ones.
We draw our conclusions and outlook in section~\ref{sec:conclu}.  

\section{$W^+W^+$ plus dijet production}
\label{sec:WW}

The process $W^+W^+ $ + 2 jets has been computed at NLO
in~\cite{Melia:2010bm}, and we fully use those results here. In this
section we recall few aspects of the calculation and refer the reader
to~\cite{Melia:2010bm} for all other details.

For the one-loop calculation one expresses one-loop amplitudes as a
linear combination of master integrals and uses $D$-dimensional
unitarity to compute the coefficients of the master integrals in this
decomposition of the amplitude. The coefficients are then given by
products of tree-level amplitudes evaluated in higher dimensions and
involving complex momenta. These tree-level helicity amplitudes are
evaluated using recursive Berends-Giele relations\cite{BG}. This is
the most natural choice since recursion relations can be easily used
to compute amplitudes involving complex momenta in an arbitrary
dimension. The master integrals are evaluated using the package
QCDloop\cite{Ellis:2007qk}.

The ingredients needed to implement a new process in the \POWHEGBOX{}
are then \cite{Alioli:2010xd}
\begin{itemize}
\item the list of the flavour structures in the Born and real
  processes for incoming and outgoing particles. Only one
  flavour structure must appear for each class of flavour structures
  that are equivalent up to a permutation of final state particles.
  In the present case we have 20 flavour structures for the Born and
  36 for the real radiation contributions;
\item the Born phase space. In our case the Born phase space involves
  an integration over 16 variables (of which one azimuthal angle is
  irrelevant);
\item the Born, real and virtual squared matrix elements. Furthermore,
  one needs the Born colour and spin-correlated amplitudes. The last one
  arises only if there are gluons as external particles, which is not the
  case in our process. The colour correlated Born amplitudes are
  available in the code of ref.~\cite{Melia:2010bm},
  where they are used in the computation of the virtual amplitudes.
  The matrix elements for each flavour
  structure should be appropriately symmetrized if identical particles
  appear in the final state;
\item the Born colour structures in the limit of large number of
  colours. Once the \POWHEG{} event kinematics and flavour structure is
  generated, we must also assign a planar colour structure to it, that is
  needed by the shower program for building the subsequent radiation and
  to model the hadronization process. In \POWHEG{} this colour assignment is
  based upon the colour structure  of the Born term in the planar
  limit\cite{Alioli:2010xd}. In the process at hand, we have two possibilities.
  In the case of two quark pairs of distinct flavour, there
  is at Born level only one diagram, therefore the leading colour
  structure is fixed. In the case of identical fermions there are two
  diagrams at Born level, corresponding to $s$ and $t$ channel
  scattering. We pick then the leading colour structure for each phase
  space point according to the value of the squared matrix element for
  $s$ and $t$ channel scattering, neglecting the interference term.
\end{itemize}

We performed the following checks on the implementation of the NLO
calculation:
\begin{itemize}
\item The \POWHEGBOX{} computes internally the soft and collinear
  limits of the real amplitude, using only the Born cross section.
  These are compared to the full real amplitude in the soft
  and collinear limits, and the results of this comparison are written
  to a file. This is a valuable check on the real and Born
  amplitudes, and is performed automatically by the \POWHEGBOX{}.
\item The \POWHEGBOX{} can also be used to compute bare LO and NLO
  distributions.  Using this feature, by fixing the same input as in
  \cite{Melia:2010bm}, we have verified that we reproduce all LO and
  NLO cross-sections and distributions presented there.
\end{itemize}
\section{Technical details in POWHEG}
\label{sec:tech}
In this section we discuss some technical issues having to do with the
\POWHEG{} implementation of the $W^+W^+$+2 jets process. We assume
that the reader has some familiarity with the \POWHEG method.

At the beginning, the \POWHEGBOX{} computes the integral of the
so called $\tilde{B}(X_i)$ function. The $X_i$, that we denote
collectively with $X$, are a set of $3 n-2$ variables (where $n$ is
the number of final state particles in the real emission process,
including decay products), defined in the unit cube, that parametrize
the momentum fraction of the incoming partons and the full phase space
for real emission. More specifically, the first $3 n - 5$ variables,
denoted collectively as $X_{\rm Born}$, parametrize the underlying
Born configuration, while the last three variables, denoted by
$X_{\rm rad}$, parametrize the radiation. The integral
\begin{equation}
\bar{B}(X_{\rm Born})=\int_0^1 d^3 X_{\rm rad}\; \tilde{B}(X)
\end{equation}
represents the inclusive cross section for the process in question at
fixed underlying Born configuration. \POWHEG{} produces events by
first generating the underlying Born configuration, and then
generating radiation using a shower technique.

The $\tilde{B}$ and $\bar{B}$ function are sum of terms, each term
referring to a specific flavour structure of the underlying Born.
\POWHEG{} first computes the integral and an upper bounding envelope
of the $\tilde{B}(X)$ function. Using this upper bounding envelope, it
is possible to generate the $X$ variables with a probability distribution
proportional to $\tilde{B}(X)$ using a hit and miss technique. The
$X_{\rm rad}$ values are discarded, and the remaining $X_{\rm Born}$
variables are thus generated with a probability proportional to
$\bar{B}$. The flavour structure of the underlying Born configuration
is chosen with a probability proportional to each flavour component of
$\tilde{B}(X)$ at the generated point.  The function $\tilde B$ itself
is the sum of the Born and the virtual contributions evaluated at the
underlying Born phase space configuration, plus an appropriate
combination of the real emission cross section, the soft and collinear
subtractions, and the collinear remnants from the subtraction of the
initial state singularities. This combination also depends upon the
$X_{\rm rad}$ variables, while the Born and virtual contributions do
not.  The evaluation of the $\tilde{B}$ function requires a
calculation of the total virtual cross section for each flavour
configuration. It turns out that one evaluation of the $\tilde B$
function requires a time of the order of 30 seconds.\footnote{This is the
  typical time on a 2.4 GHz CPU, if the program is compiled with the
  {\tt ifort} compiler.}  Although this seems to be a fairly long
time, as long as the problem can be trivially parallelized on a large
CPU cluster, it can be dealt with. In fact, however, the problem can
be parallelized only after the importance sampling integration grid
has been established. It is thus common practice, in this kind of NLO
calculations, to build the adaptive integration grid using only the
Born contribution.  In our case, we introduced a switch in our input
file, called {\tt fakevirt}. If this token is set to one, the virtual
correction is replaced by a term proportional to the Born cross
section. We thus perform the first integration step, when the adaptive
integration grid is formed, with this token set, so that no calls to
the virtual routines are performed.  This way, it is not difficult to
obtain reasonably looking adaptive grids with 500000 calls to
the $\tilde{B}$ function, taking
about 10 hours of CPU time. The same calculation using the full
virtual contribution would take a time of the order of 170 days, and would
thus be unfeasible.

After the importance sampling grid has been established, the
computation of the integral of the $\tilde{B}$ function, and the
computation of the upper bounding envelope that is used for the
generation of the underlying Born configurations, can be performed in
parallel.  The \POWHEGBOX{} already had a mechanism to perform this
stage of the computation in parallel and to combine all the results.

We have found that it is convenient to use the
so called ``folding'' technique in the integration of the $\tilde{B}$
function. The folding procedure is better explained by an
example. Given a function $f(x)$ to be integrated by a Monte Carlo
technique in the range $0<x<1$, one can replace it by the function
\begin{equation}
F(y)=\frac{1}{m}\sum_{i=0}^{m-1} f\left(\frac{i+y}{m}\right),
\end{equation}
also defined for $0<y<1$. It is obvious that
\begin{equation}
\int_0^1 F(y)\; d y=\int_0^1 f(x)\; d x\,.
\end{equation}
It is also obvious that the larger $m$ is, the smoother the $F$
function will be, thus requiring less points in a Monte Carlo
integration. We call this procedure ``folding'' the $x$ variable.  In
the \POWHEGBOX{}, the radiation variables can be folded
individually.\footnote{In fact, rather than the $X_{\rm rad}$
  variables, what is folded are the corresponding variables, piecewise
  linear functions of the $X_{\rm rad}$, that have constant importance
  sampling in the adaptive grid.}  In previous works, the use of
folding was advocated to avoid spurious negative weights. In the
present case, besides also serving that purpose, folding is used to
balance the computer time needed for the computation of the real and
virtual contribution. In fact, since only the radiation variables are
folded, the virtual contribution is the same in a given folded set,
and thus is computed only once.  The real contribution is instead
computed several times. By using this procedure, the $\tilde{B}$
function becomes a smoother function of the radiation variables, so
that the integration becomes easier to perform, and also the
generation of the underlying Born configurations becomes more
efficient.  It is found that with folding numbers of 5, 5, 10,
referring respectively to the radiation variables $\xi$, $y$ and
$\phi$, the time required to compute the virtual contribution becomes
comparable to the time for the evaluation of the real one.  As we said
earlier, by using the Intel Fortran compiler, the time for a single
call to the virtual cross section is roughly 30 seconds. In order to
get 500000 points, one needs about 170 days of computer time, a
relatively easy task on modern days clusters with hundreds of CPU's. A
further problem arises, however.  Assuming that we are using 100
CPU's, each process generates 5000 points. This is not enough 
to get a reasonable upper bounding envelope for the generation of the
underlying Born configuration.  In fact, the procedure used in the
\POWHEGBOX{} (described in ref.~\cite{mint}) has no time to reach the
upper bound with such a small number of points.  Even when combining
together the upper bounds of the different runs, one gets an
unacceptable rate of upper bound failures in the generation of the
underlying Born configuration, of the order of 1 every ten calls to
the $\tilde{B}$ function, which can thus affect final distributions.
We modified the \POWHEGBOX{} basic code, in order to deal with
this problem. In short, the return values of the $\tilde{B}$ function
calls were first written to files by the parallel processes, and the
upper bounding envelope was later evaluated by reading all the
generated files. After this step, the program is capable of generating
user process events (that is to say, events ready to be fed
through a shower Monte Carlo program).  The efficiency, however, turns out
to be very small, of the order of 2\%{}. This means that the
generation time is of the order of $30/0.02=1500$ seconds, roughly a
couple of events per hour. In order to collect 100000 events, we would
thus need 500 hours on a typical 100 CPU's cluster. We were able to
reach 15\% efficiency by a further technical modification to the
\POWHEGBOX{} code which is described in appendix~\ref{Technicalities}.

At this point, no further problems arise. One can generate the upper
bound normalization for the generation of radiation,
and start the event generation using a
parallel CPU cluster. The upper bound failures in the generation of
the underlying Born configurations are at the level of 2 for 1000
generated events. Due to the large folding numbers, the fraction of
negative weighted events is only 0.4\%{}, an acceptable value.
The generation time is about three minutes per
event. Most of the generation time is consumed by the computation of
$\tilde{B}$. Since the efficiency in the generation of the underlying
Born is of the order of 10\%, the generation time is of the order of
several times the time needed for the computation of a single virtual
point. Again, having a large CPU cluster at one's disposal, it is not
hard to generate few hundred thousands events.

It is also worth asking whether other improvements in performance are
actually possible.  Aside from considering more aggressive hardware
requirements, like using GPU's and the like, we have immediately noted
another speed aspect that can be improved by modifying suitably the
\POWHEGBOX{} code.  In fact, at the moment, we compute the full
$\tilde{B}$ function when we generate the underlying Born kinematics,
and decide its flavour structure on the basis of the size of each flavour
contribution to it. This is what was implemented in the \POWHEGBOX{},
mainly for reasons of simplicity. A more efficient approach would be
to store sufficient information to generate each underlying Born
flavour configuration individually. In this way, the generation
process would start by picking an underlying Born flavour
configuration with probability proportional to the corresponding
contribution to the total cross section. Given the underlying Born
flavour configuration, one would then generate the underlying Born
phase space. It is not unlikely that, with this approach, one may gain
a factor of order 10 in speed.

Alternatively, a speed gain may be achieved if the code that computes
the virtual contribution is optimized to compute all flavour structure
contributions to the virtual cross section at once.

\section{Results}
\label{sec:res}

In this section we present our results. 
We consider proton-proton collisions with center-of-mass energy
$\sqrt{s} = 7~ {\rm TeV}$. We require the $W^+$ bosons to decay
leptonically into $e^+ \mu^+ \nu_e \nu_\mu $. The $W$-bosons are
produced on mass-shell and we assume a diagonal CKM matrix.
Neglecting interference effects, which are numerically suppressed since
they force the $W$ bosons off mass-shell, the cross-section for
same-flavour production is half that of different flavour. This implies that
the full cross-section summing over electrons and muons can be
obtained by multiplying the results presented here by a factor two.

The setup used is largely inspired by~\cite{Melia:2010bm}, but we
consider here a different center of mass energy. In the distributions
shown, we impose the following leptonic cuts
\begin{equation}
p_{t, l^+} > 20\; {\rm GeV}, \qquad 
p_{t, \rm miss} > 30\; {\rm GeV}, \qquad 
| \eta_{l^+}| < 2.4\;. 
\label{eq:lepcuts}
\end{equation}
We define jets using the anti-$k_{\perp}$
algorithm~\cite{Cacciari:2008gp}, with the $R$ parameter set to 0.4.
We do not impose any transverse-momentum cut on the two outgoing jets,
nor do we impose lepton isolation cuts.

The mass of the $W$-boson is taken to be $m_W = 80.419~{\rm GeV}$, the
width $\Gamma_W = 2.141$~{\rm GeV}. $W$ couplings to fermions are
obtained from $\alpha_{\rm QED} (m_Z) = 1 / 128.802$ and $\sin^2
\theta_W = 0.2222$.  We use MSTW08NLO parton distribution functions,
corresponding to $\alpha_s(M_Z) = 0.12018$~\cite{Martin:2009iq}. We
consider the top quark infinitely heavy and neglect its effects, while all
other quarks are treated as massless.  We set the factorization scale equal
to the renormalization scale, which we choose to be 
\begin{equation}
\mu_R = \mu_F =
\frac{p_{t, p1} +p_{t, p2} +E_{t, W_1} +E_{t, W_2}}{2}\,,
\qquad 
E_{t, W} = \sqrt{M_W^2+p_{t,W}^2}\,, 
\end{equation}
where  $p_{t,W_1}$  $p_{t,W_2}$, $p_{t, p1}$ and $p_{t, p2}$ and are the
transverse momenta of the two $W^+$ and of the two emitted partons in the
underlying Born configuration. 
We use \PYTHIA{} 6.4.21~\cite{Sjostrand:2006za} to shower the events,
include hadronization corrections and underlying event effects, with
the Perugia 0 tune (i.e. we call {\tt PYTUNE(320)} before calling {\tt
  PYINIT}).  We have also showered the events using \HERWIG{}
\cite{Corcella:2000bw}.  We found only marginal differences between
the \HERWIG{} and \PYTHIA{} results. Thus, we do not show any plot of
\HERWIG{} results.

We remind the reader that we consider here only the QCD production of
$W^+ W^+ jj$, while we completely neglect the electroweak production.
We also stress that we have not computed any theoretical error due to
scale variations or PDF uncertainties on our distributions. Thus, our
error bars are only statistical. The purpose of the plots we show is
only to validate our results. Since the code is public, a user
may study theoretical uncertainties at will.

With the setup described above, we obtain an NLO total cross-section of
$2.74 \pm 0.03$ fb, that coincides by construction with the
\POWHEGpPYTHIA{} result.  If we impose the leptonic cuts of
eq.~(\ref{eq:lepcuts}), and do not require any minimum transverse
momentum for the jets, we have a NLO cross section of $1.11\pm 0.01$
fb, and a slightly lower cross section with \POWHEGpPYTHIA{} of $1.06
\pm 0.01$ fb.  Unless otherwise stated, these are the cuts applied to
the distributions presented in the following.
If we also require to have at least two jets with transverse momentum
larger than $30$ GeV we obtain an NLO inclusive cross-section of
$0.84 \pm 0.01$ fb. A $30$~GeV transverse momentum cut was also
applied to the third hardest jet, when we plot its rapidity
distribution.
\FIGURE[t]{
\includegraphics[angle=0,scale=0.58]{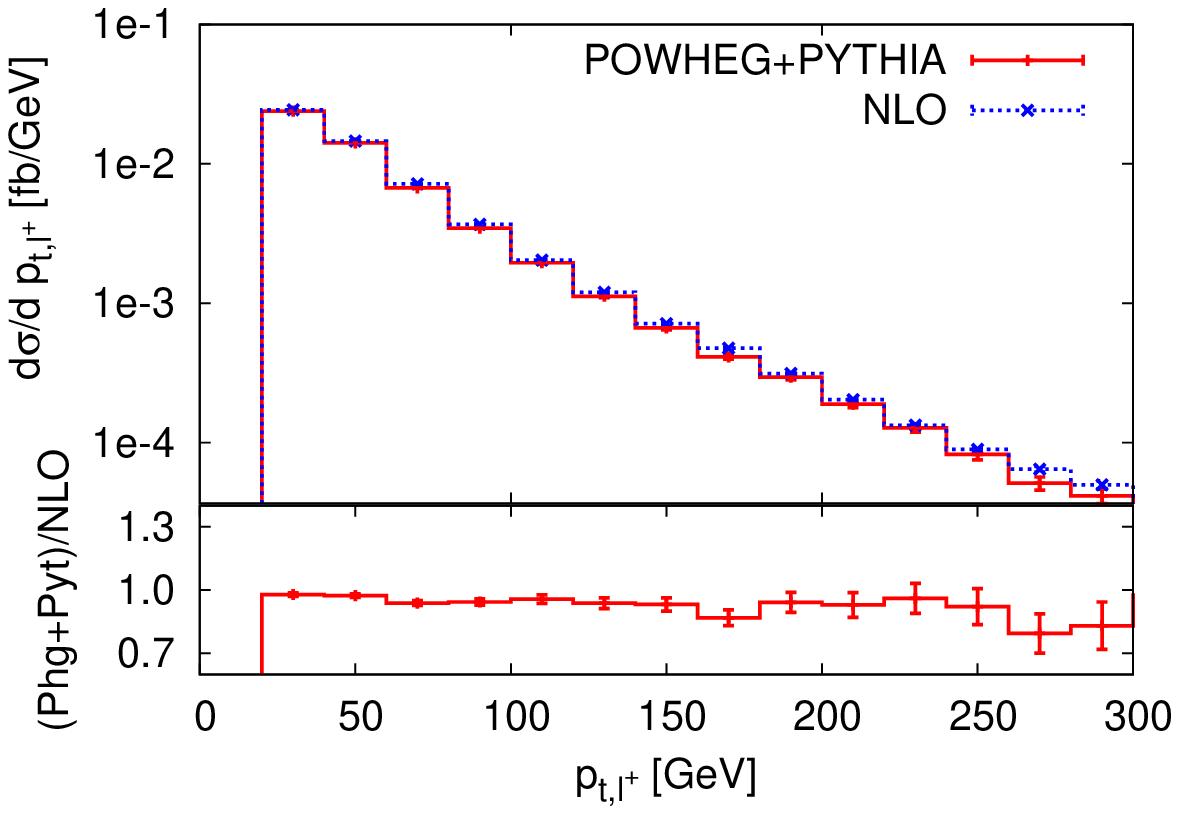}
\includegraphics[angle=0,scale=0.58]{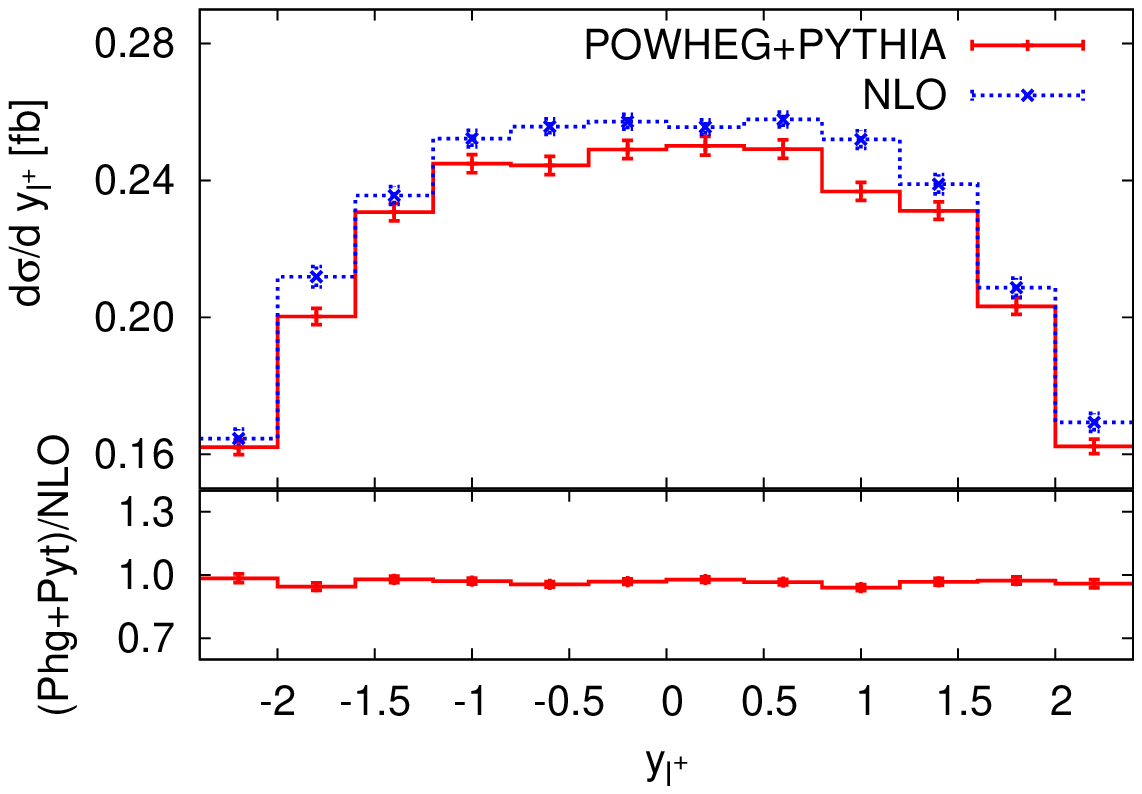}
\vspace{0.4cm}
\includegraphics[angle=0,scale=0.58]{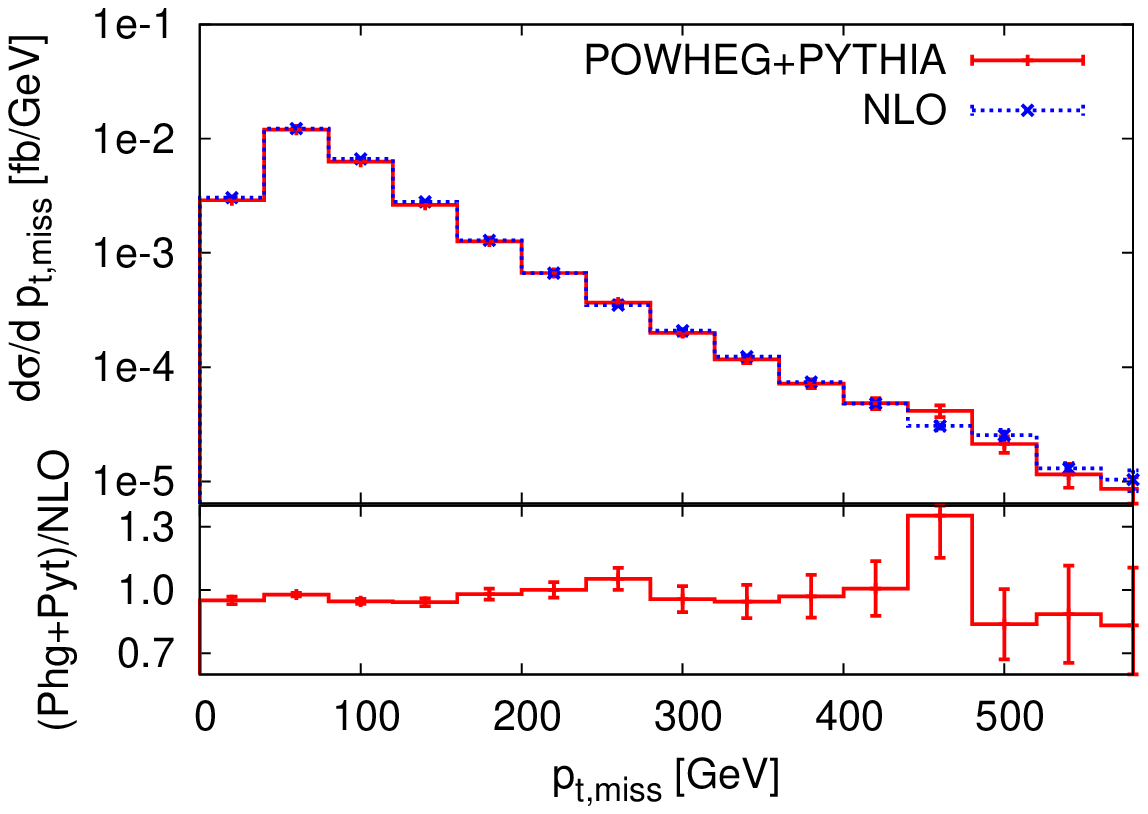}
\includegraphics[angle=0,scale=0.58]{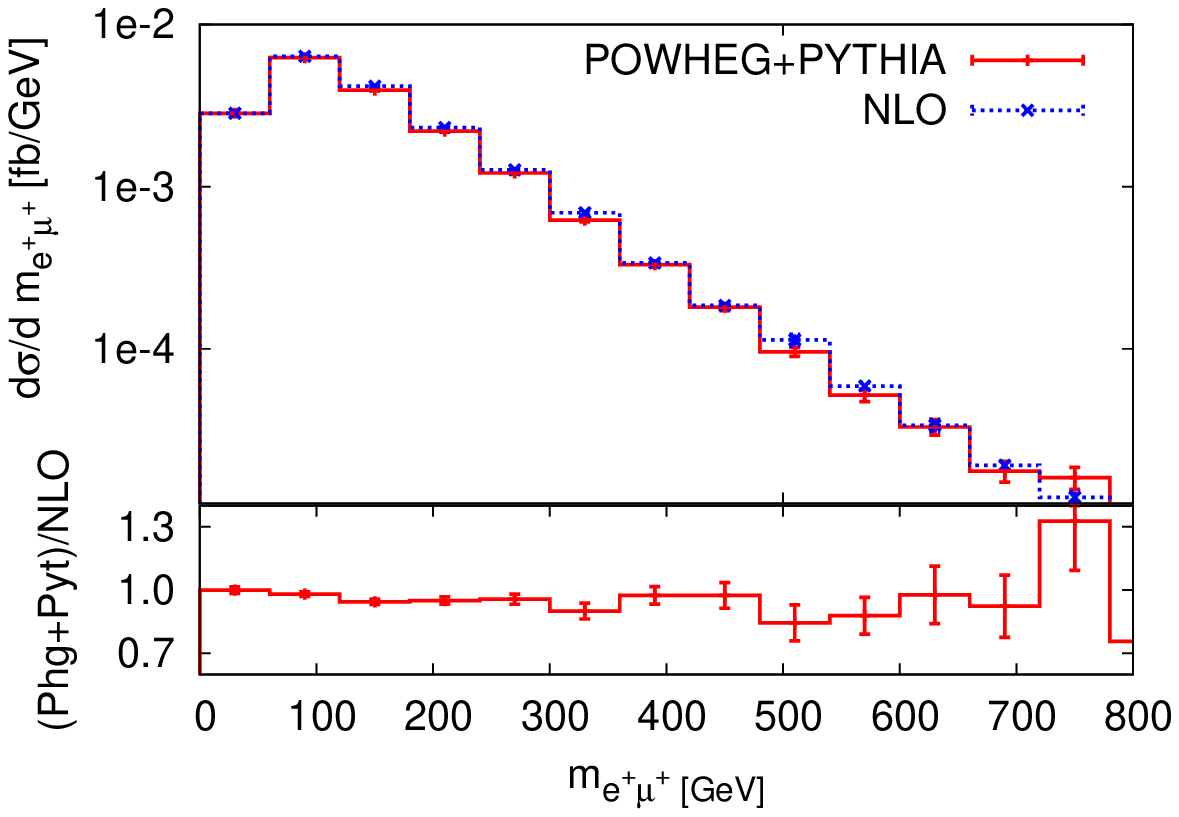}
\vspace{0.4cm}
\includegraphics[angle=0,scale=0.58]{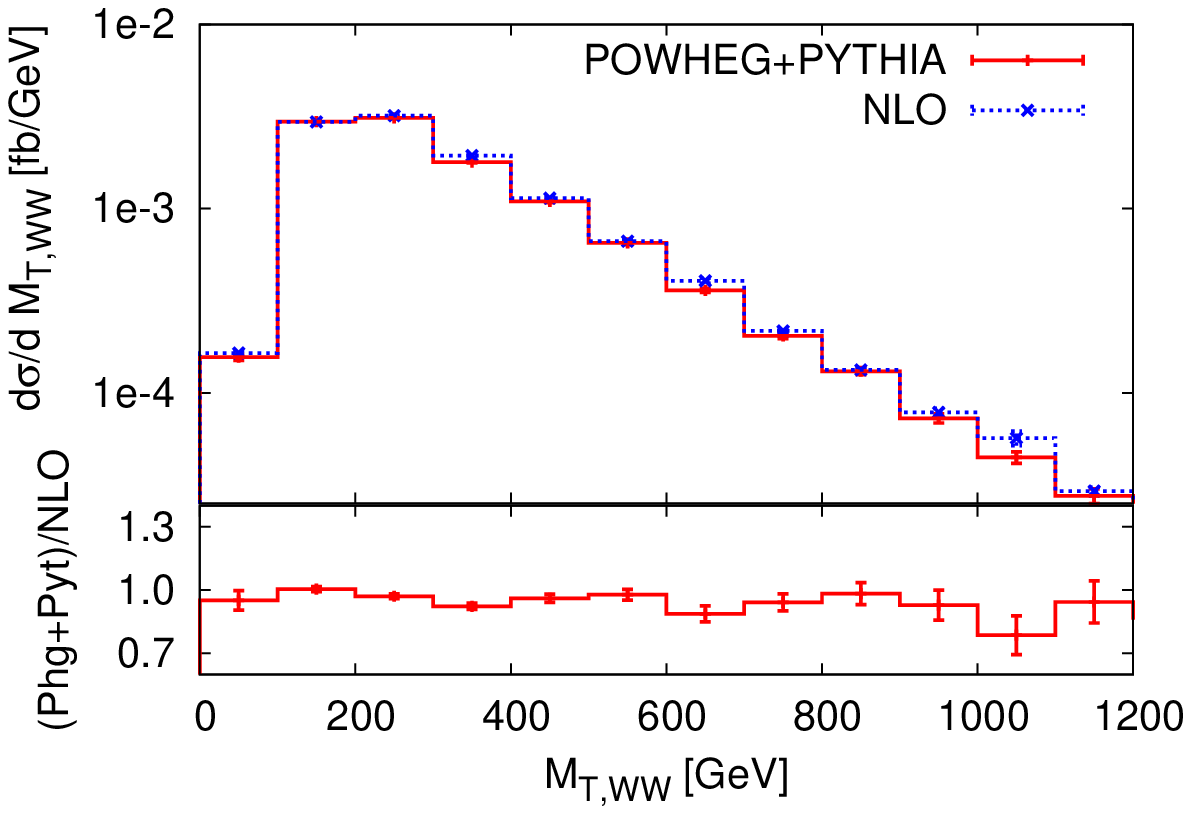}
\caption{Leptonic kinematic distributions for the QCD production of
  $pp \to e^+\, \mu^+\, {\nu}_{e}\, {\nu}_{\mu}\, + 2~{\rm jets}$ at
  next-to-leading order and with \POWHEGpPYTHIA{}. See text for more
  details.}
\label{fig1}
}

We now discuss some kinematical distributions.  We first consider
leptonic inclusive distributions.  We plot in fig.~\ref{fig1} the
inclusive transverse momentum distribution for the charged lepton
($e^+$ or $\mu^+$) $p_{t, l^+}$ and its rapidity distribution
$y_{l^+}$, the missing transverse momentum, the charged lepton system
invariant mass $m_{e^+\mu^+}$, and the transverse mass of the two $W$
bosons defined as
\begin{equation}
m_{\rm {T,WW}}^2 = (E_{T, e^+ \mu^+} +
\tilde{E}_{T, \rm miss})^2 - ( {\bf p}_{t, e^+ \mu^+} + {\bf
  p}_{t, \rm miss})^2\,,
\end{equation}
where the missing transverse energy $\tilde{E}_{T, \rm miss}$ is
reconstructed from the missing transverse momentum using the invariant
mass of the charged lepton system $\tilde{E}_{T, \rm miss} =
\sqrt{{\bf p}_{t, \rm miss}^2 + m_{e^+ \mu^+}^2}$.  For these
distributions we find good agreement between NLO and \POWHEGpPYTHIA{},
and do not observe any relevant difference in shape.

\FIGURE[t]{
\includegraphics[angle=0,scale=0.55]{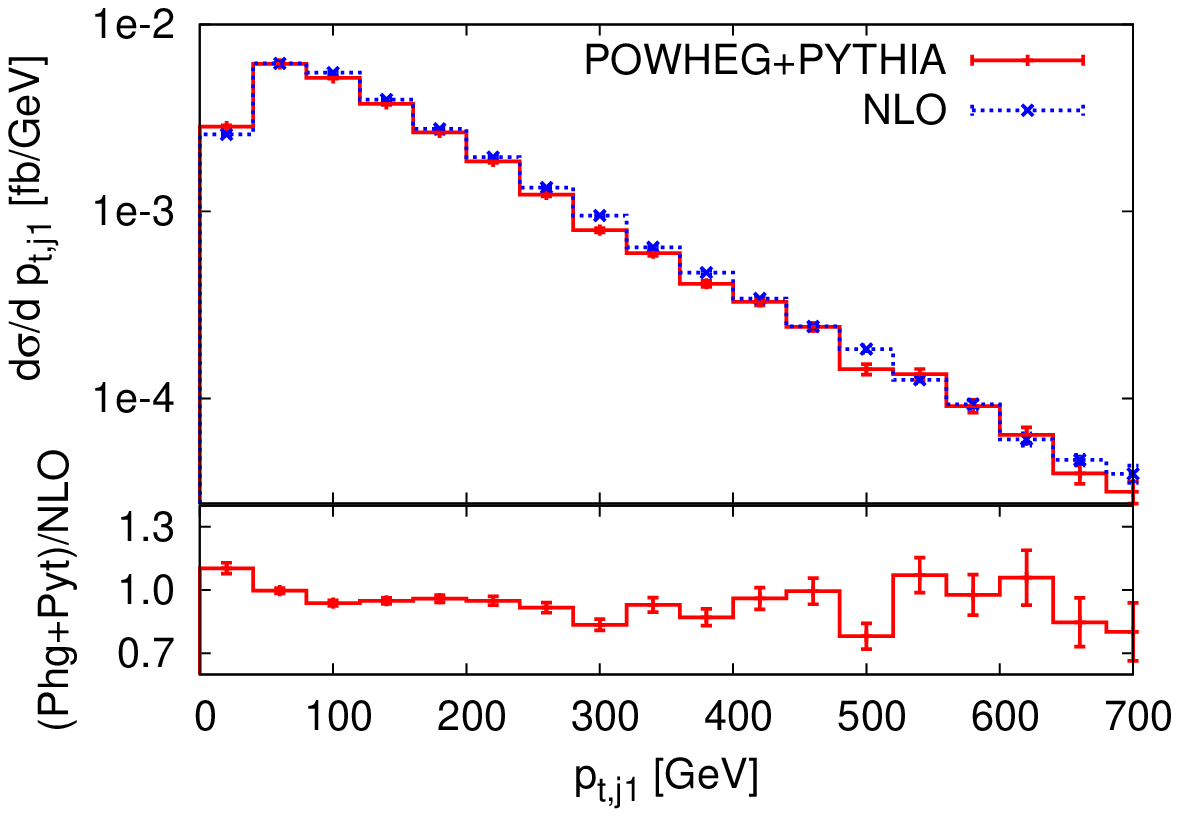}
\includegraphics[angle=0,scale=0.55]{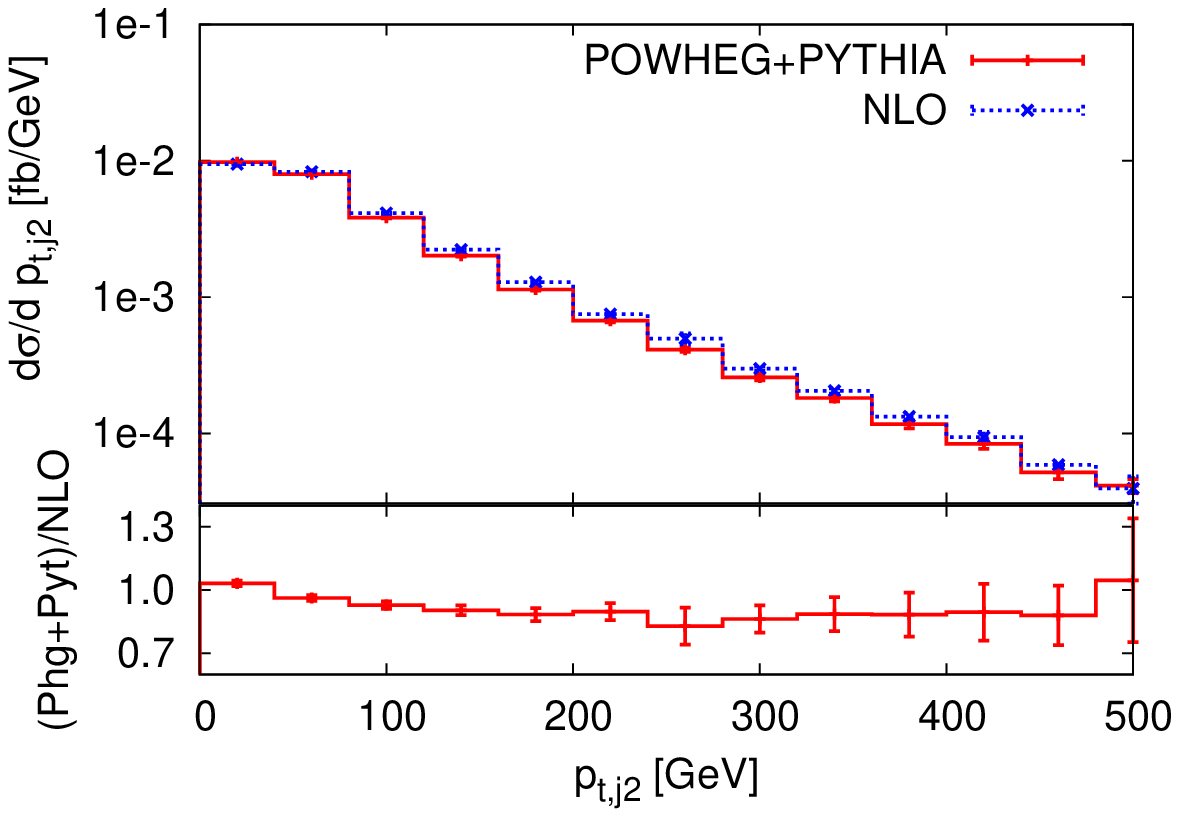}

\vspace{0.4cm}

\includegraphics[angle=0,scale=0.55]{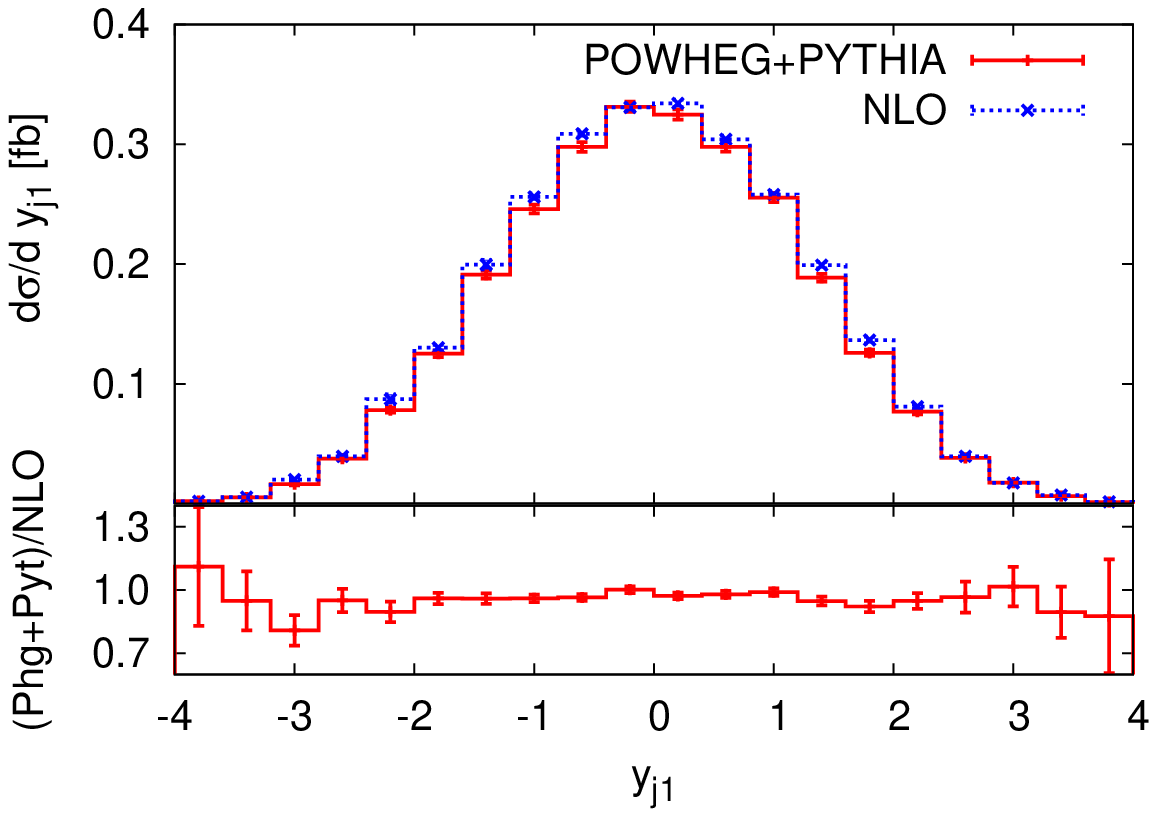}
\includegraphics[angle=0,scale=0.55]{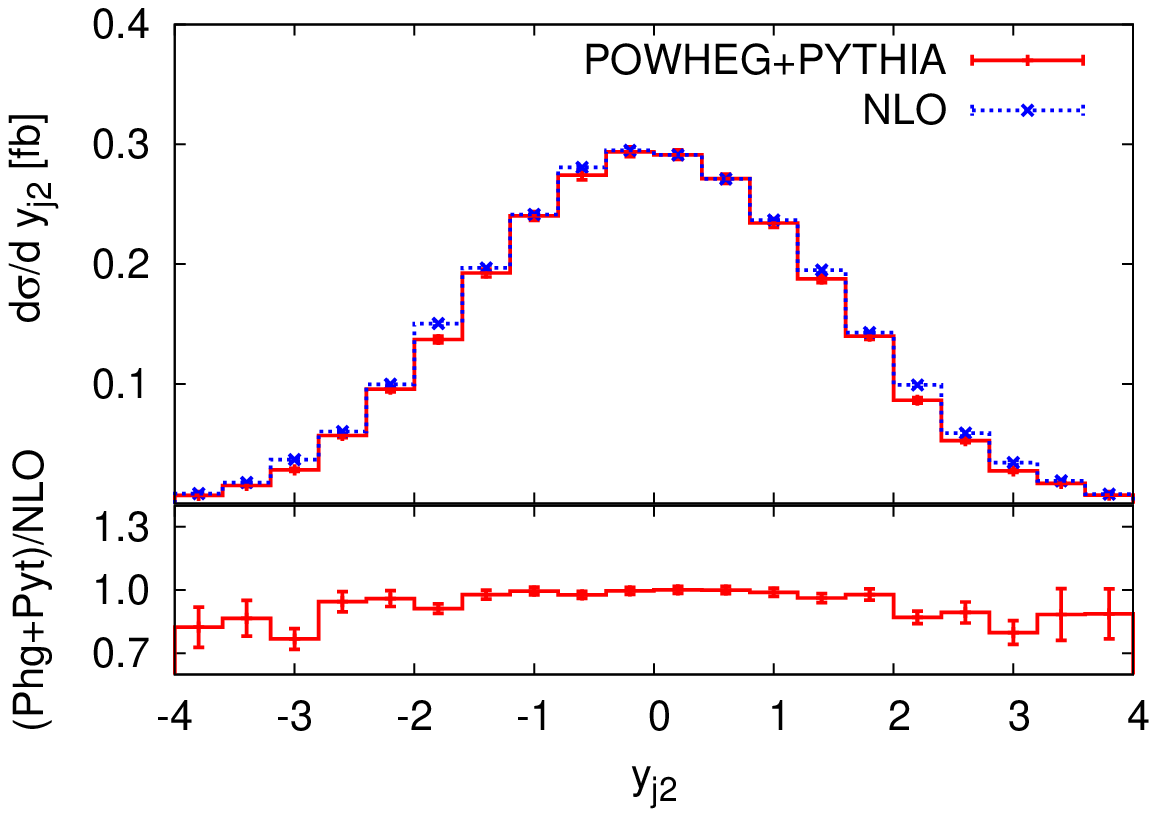}

\vspace{0.4cm}

\includegraphics[angle=0,scale=0.55]{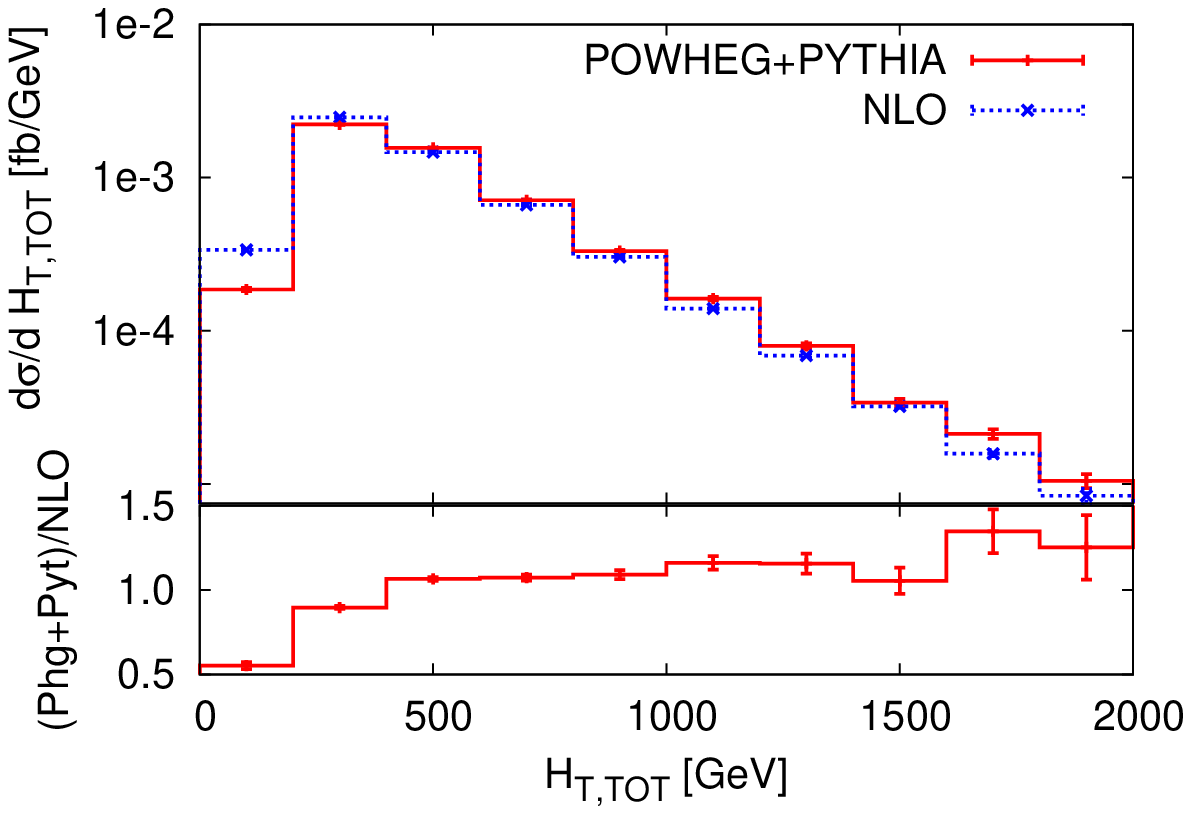}
\caption{Hadronic kinematic distributions for the QCD production of
  $pp \to e^+\, \mu^+\, {\nu}_{e}\, {\nu}_{\mu}\, + 2~{\rm jets}$ at
  next-to-leading order and with \POWHEGpPYTHIA{}. See text for more
  details.}
\label{fig2}
}

In fig.~\ref{fig2} we show some hadronic inclusive distributions.  We
plot the transverse momentum and rapidity distributions of the two
leading jets, i.e. those with largest transverse momentum, and the
total transverse energy of the event $H_{\rm T,TOT}$, defined as
\begin{equation}
H_{\rm T,TOT} =
p_{\rm t, e^+}+ p_{\rm t, \mu^+}+ p_{\rm t, miss}+ \sum_{\rm j} p_{\rm t,
  j}\,, 
\end{equation}
where the sum runs over {\it all} jets in the event.  For the
transverse momentum and rapidity distributions we notice differences
of the order of 10\%{} between the NLO and \POWHEGpPYTHIA{}
results. We also find that the \POWHEGpPYTHIA{} distributions tend to
be more peaked for smaller jet transverse momenta, and also that jets
tend to be slightly more central.

The $H_{\rm T,TOT}$ distribution, on the other hand, displays large
differences, especially on the first bin, where the \POWHEGpPYTHIA{}
result is a factor 2 smaller that the NLO one.  This feature is easily
explained. The shower and the underlying event generated by \PYTHIA{}
adds several soft particles to the event. Since we do not apply any
transverse momentum cut, these soft particles are clustered in jets,
and contribute positively to $H_{\rm T,TOT}$. Because of the large
multiplicity of LHC events, assuming a typical transverse momentum of
500~MeV for soft hadrons, we see that it is not inconceivable that
this increase may reach values of the order of 50~GeV. The first bin
of the distribution is the most affected one, because this mechanism
can cause events to migrate to higher bins from it, while no event
will migrate backward.  This explanation is easily tested. First of
all, we show in the left plot of fig.~\ref{httotmore} that the
\POWHEG{} user process event, without \PYTHIA{} shower, is in good
agreement with the NLO result.
\FIGURE[t]{
\includegraphics[angle=0,scale=0.55]{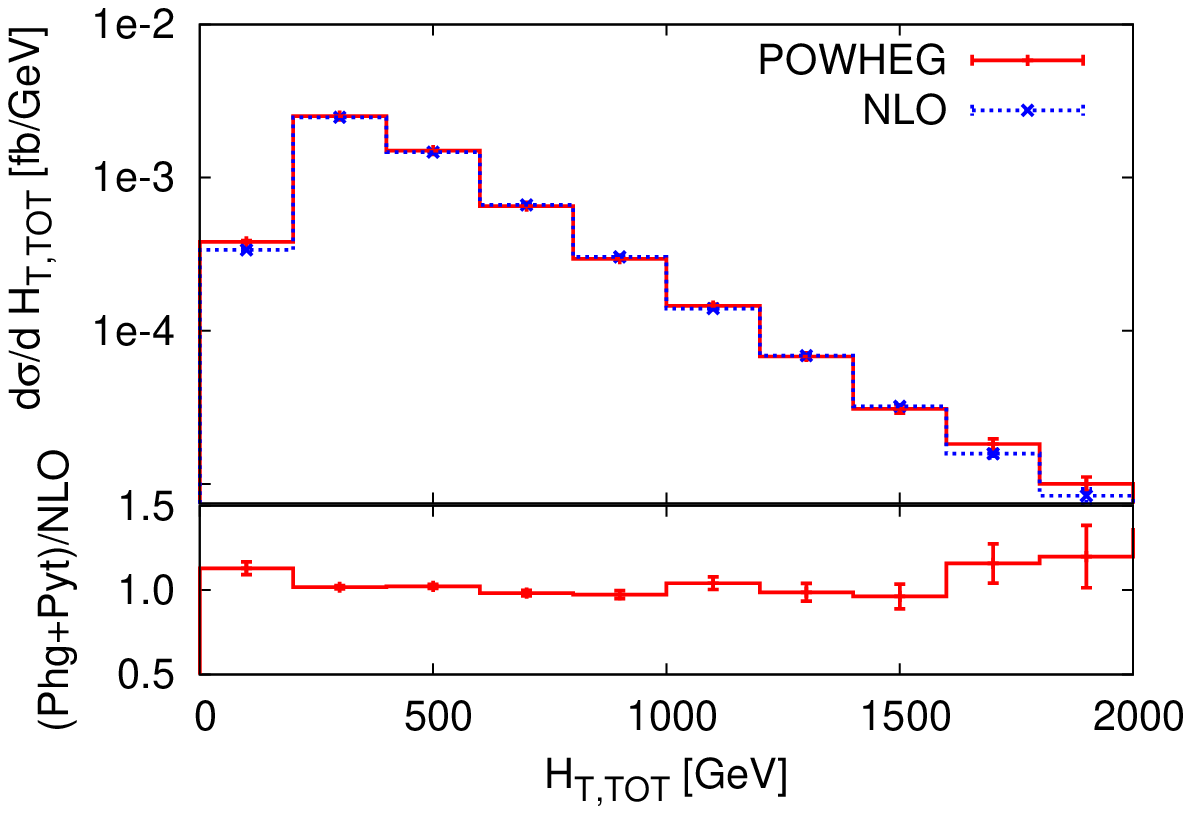}
\includegraphics[angle=0,scale=0.55]{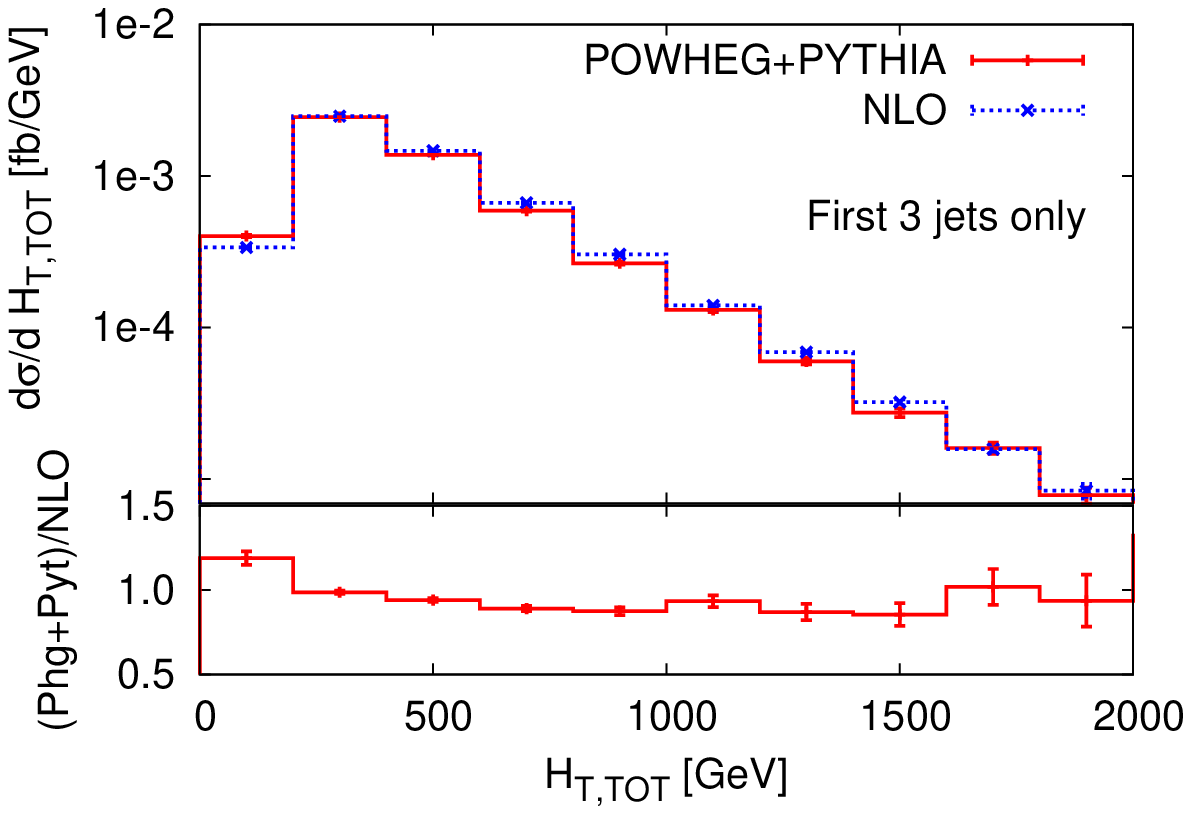}
\vspace{0.4cm}
\caption{On the left we show a comparison of the NLO and bare
\POWHEG{} distribution. On the right we show the \POWHEGpPYTHIA{}
and the NLO $H_{\rm T,TOT}$ distribution when only the three hardest
jets are considered in the computation.}
\label{httotmore}
}
We see that, if anything, it is the \POWHEG{} distribution that is slightly
above the NLO one.  This proves that this feature is not originated by
the \POWHEG{} implementation. In the right plot, we show a comparison
of the \POWHEGpPYTHIA{} and the NLO calculation for the $H_{\rm
  T,TOT}$ distribution, this time defined to involve only the three
hardest jets.  The NLO distribution, of course, is not affected. On the
other hand, the \POWHEGpPYTHIA{} is brought in much better agreement
with the NLO calculation.

\FIGURE[t]{

\includegraphics[angle=0,scale=0.55]{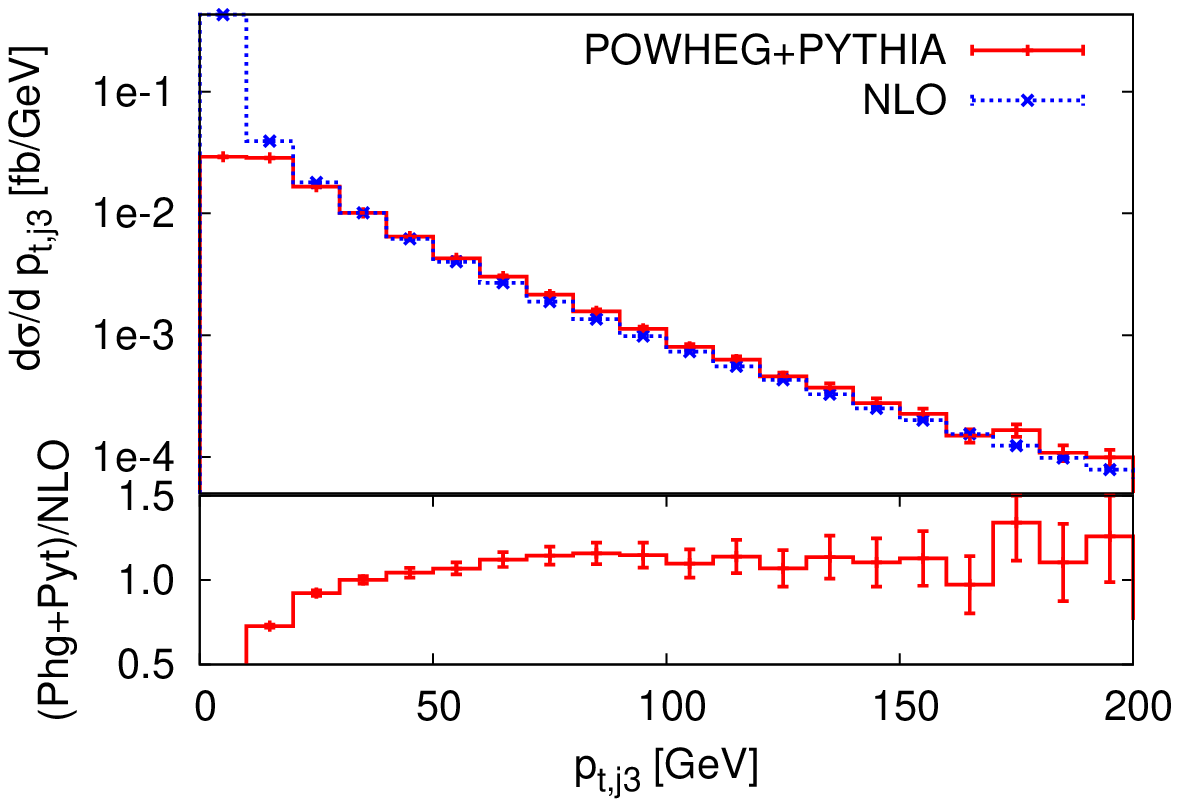}
\includegraphics[angle=0,scale=0.55]{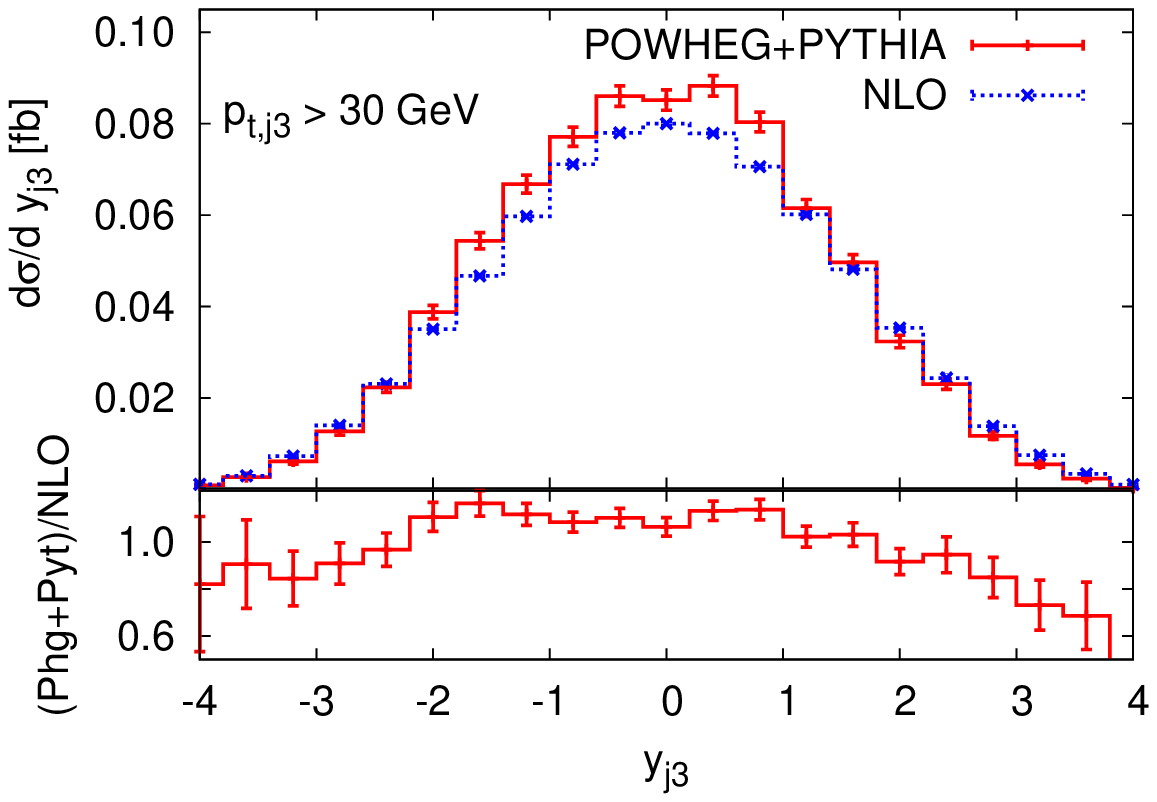}

\vspace{0.4cm}

\includegraphics[angle=0,scale=0.55]{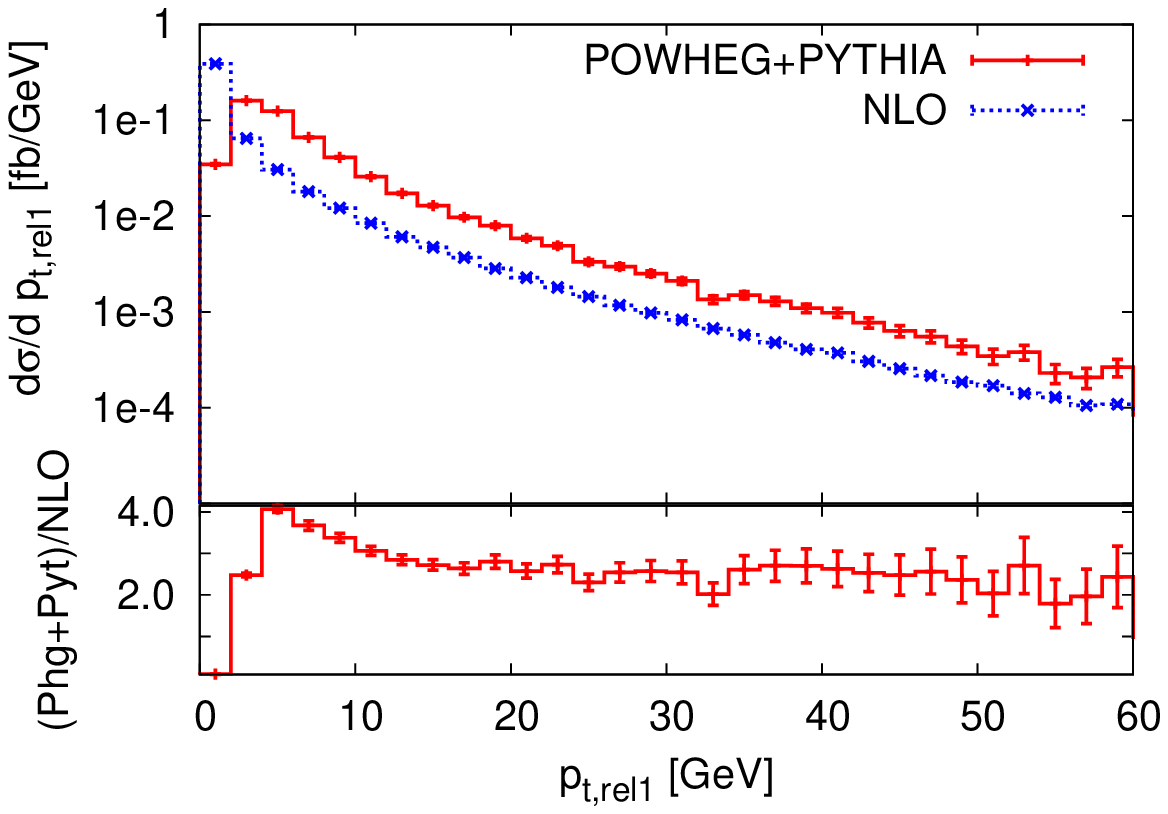}
\includegraphics[angle=0,scale=0.55]{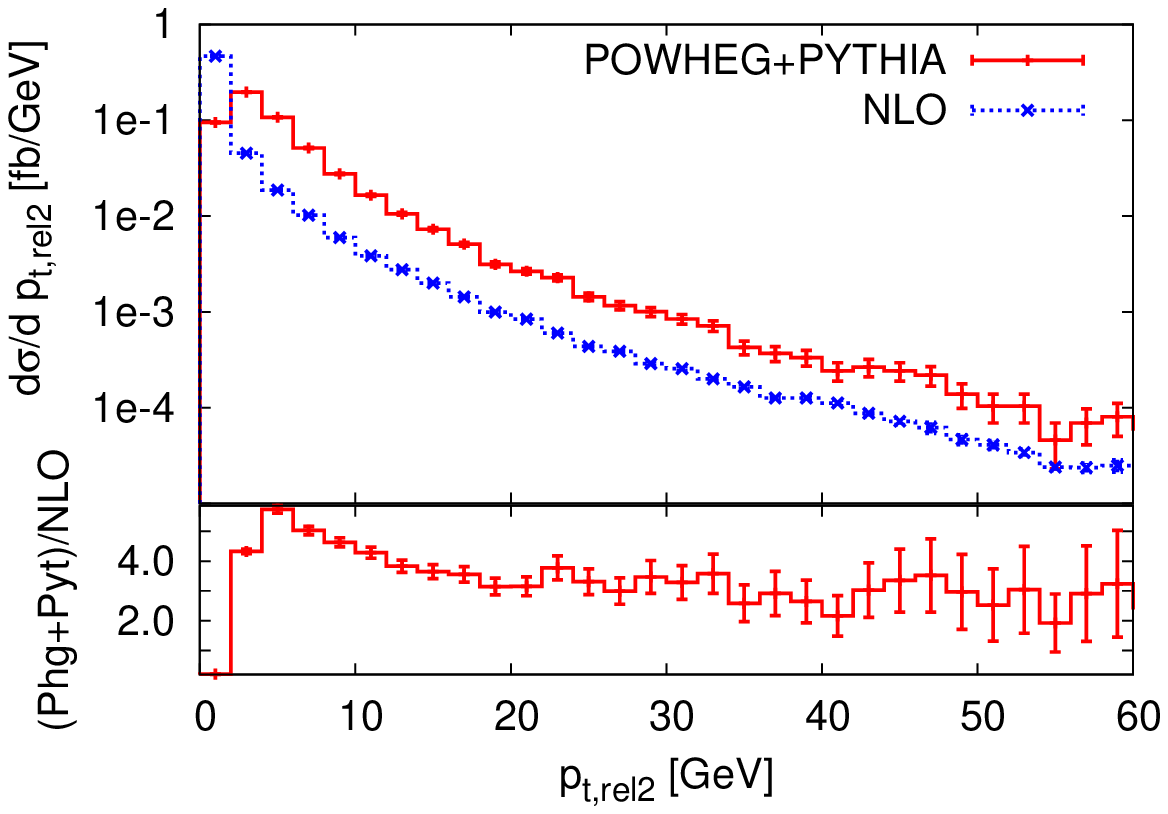}

\caption{Hadronic kinematic distributions for the QCD production of
  $pp \to e^+\, \mu^+\, {\nu}_{e}\, {\nu}_{\mu}\, + 2~{\rm jets}$ at
  next-to-leading order and with \POWHEGpPYTHIA{}. See text for more
  details.}
\label{fig3}
}

Finally, we show in fig.~\ref{fig3} the transverse momentum and
rapidity of the third jet (in the last distribution we impose a
transverse momentum cut of 30 GeV on the jets), and the relative
transverse momentum distribution of the particles inside the two
leading jets defined with respect to the jet axis in the frame where
the jet has zero rapidity
\begin{equation}
p_{\rm t, rel j} = \sum_{i \in j}\frac{|\vec k_i \wedge \vec p_j|}{|\vec p_j|}\,. 
\label{eq:ptrel}
\end{equation}
Here $k_i$ denotes the momentum of the $i^{\rm th}$ particle and $p_j$
of the $j^{\rm th}$ jet.
At NLO it is only the real radiation that contributes to these four
distributions, and we see clearly a divergence at small $p_{t, j3}$
and $p_{\rm t rel, j}$, while in the \POWHEGpPYTHIA{} prediction the
distribution has a Sudakov peak and goes to zero for $p_{t, j3},
p_{\rm t rel, j} \to 0 $. We also remark that the third jet tends to
be more central with \POWHEGpPYTHIA{}.

\section{Conclusions}
In this work we have presented a \POWHEG{} implementation for the QCD
production of $p p\to W^+ W^+$ plus two jets, at NLO in the strong
coupling constant, with $W$ leptonic decays included with NLO accurate
spin correlations. The NLO corrections for this process have been
computed recently using $D$-dimensional unitarity in
ref. \cite{Melia:2010bm}.  In this work we just focused upon building
up a \POWHEG{} implementation, in order to consistently interface the
calculation to shower Monte Carlo generators. The \POWHEG{}
implementation was built in the framework of the
\POWHEGBOX{}~\cite{Alioli:2010xd}.

The $p p\to W^+ W^+ 2j$ process is of considerable phenomenological
interest, being an important background to new physics signatures, and
to the study of double parton scattering phenomena.  Furthermore, its
study is also interesting since it represents a first \POWHEG{}
implementation of a complex, $2 \to 4$ scattering process, where the
calculation of the virtual corrections is highly demanding from a
computational point of view. Besides this issue, the \POWHEG{}
implementation of this process does not present any special
problem. The Born cross section is finite, in spite of the presence of
the two jets in the final state, so, from this point of view the
process is similar to Higgs boson
production in Vector Boson Fusion~\cite{Nason:2009ai}.
However, the large amount of computer
time required for the calculation of the virtual contributions has in
practice turned out into a difficult problem to deal with, so that the
\POWHEGBOX{} implementation of the process was in fact not completely
trivial.

We have spotted a number of possible improvements to the \POWHEGBOX{}
code that can result in a substantial increase in efficiency, and we
have implemented the most simple ones. We were thus able to
generate an adequate number of events for this process and, most
importantly, we have convinced ourselves that the \POWHEGBOX{}
efficiency can be increased even further, in order to match the level
of complexity that is now possible in NLO calculations.

We have compared the \POWHEGBOX{} result, interfaced with the
\PYTHIA{} and \HERWIG{} Monte Carlo, with the bare NLO one, and have
found consistency with the features observed in other implementations:
very inclusive observables, like the lepton spectra, display a
remarkable agreement; quantities involving leading jets also agree
well, with only minor differences; quantities involving the radiated
jet display marked differences in the Sudakov region.

Finally, we have made our code public. It can be retrieved by
following the instructions at the \POWHEGBOX{} web site
\url{http://powhegbox.mib.infn.it}.

\label{sec:conclu}
\vskip 0.5cm
{\bf Acknowledgments} 
We thank Kirill Melnikov, Carlo Oleari and Emanuele Re for useful exchanges.
The calculations have been performed using TURING, the INFN computer cluster
in Milano-Bicocca, and the centralized INFN CSN4Cluster.  This work is
supported by the British Science and Technology Facilities Council and
by the LHCPhenoNet network under the Grant Agreement
PITN-GA-2010-264564.

\appendix
\section{Raising the generation efficiency}\label{Technicalities}
Due to the large amount of computer time needed to compute virtual
corrections, it is mandatory to increase the generation efficiency for
the underlying Born configurations. For processes with many external
legs, this efficiency can be in fact quite low.

The \POWHEGBOX{} generates the underlying Born and radiation
kinematics using a hit and miss technique. An upper bounding envelope
of the $\tilde{B}$ function is found, of the form
\begin{equation}\label{eq:mintub}
\tilde{B}(X) \le \prod_{i=1}^n f^{(i)}(X_i),
\end{equation}
where the $X_i$ are the integration variables, and the $f^{(i)}$
functions are step functions of the integration variables. The size of
the step is determined by the importance sampling grid itself, as
documented in ref.~\cite{mint}. In order to generate a configuration,
the points $X_i$ are first generated with a probability distribution
equal to $f^{(i)}(X_i)$.  Then a uniform random number $r$, with
\begin{equation}
0\le r \le \prod_{i=1}^n f^{(i)}(X_i)
\end{equation}
is generated. One then computes $\tilde{B}(X)$. If $r\le \tilde{B}(X)$
we have a hit, and the configuration is kept. Otherwise the
configuration is rejected (we have a miss), and we restart the
procedure.

It is clear that, if the number of integration variables is large (as
in our case), an upper bound of the form eq.~(\ref{eq:mintub}) will
generally be highly inefficient, just because the product of a large
number of terms will tend to build up large values.  In order to
remedy to this problem, we have exploited the fact that the
$\tilde{B}$ function is equal to the Born cross section plus higher
order terms. It is thus natural to expect that an upper bound of the
form
\begin{equation}\label{eq:tighterbound}
\tilde{B}(X) \le B(X)\times\prod_{i=1}^n g^{(i)}(X_i),
\end{equation}
will be much closer to it. We thus determine the $g$ functions for
this bound, using the same technique used for the $f$ functions.
Then we modified our code in such a way
that, before computing $\tilde{B}$ in order to test for a hit or miss,
we compute the right hand side of eq.~(\ref{eq:tighterbound}). If it is
smaller than $r$, $\tilde{B}$ will also be smaller than $r$, and we
thus know that we have a miss without the need to compute the time
consuming $\tilde{B}$ function.  By adopting this method, we have
reached an efficiency of 15\%{}, instead of a 1-2\%{} efficiency that
we achieve with the \POWHEGBOX{} default method.


\begin{thebibliography}{99}

\bibitem{Bredenstein:2009aj}
  A.~Bredenstein, A.~Denner, S.~Dittmaier, S.~Pozzorini,
  %``NLO QCD corrections to pp ---> t anti-t b anti-b + X at the LHC,''
  Phys.\ Rev.\ Lett.\  {\bf 103 } (2009)  012002.
  [arXiv:0905.0110 [hep-ph]].
  
\bibitem{Bredenstein:2010rs}
  A.~Bredenstein, A.~Denner, S.~Dittmaier, S.~Pozzorini,
  %``NLO QCD corrections to top anti-top bottom anti-bottom production at the LHC: 2. full hadronic results,''
  JHEP {\bf 1003 } (2010)  021.
  [arXiv:1001.4006 [hep-ph]].

\bibitem{Bevilacqua:2009zn}
  G.~Bevilacqua, M.~Czakon, C.~G.~Papadopoulos, R.~Pittau, M.~Worek,
  %``Assault on the NLO Wishlist: pp ---> t anti-t b anti-b,''
  JHEP {\bf 0909 } (2009)  109.
  [arXiv:0907.4723 [hep-ph]].
  
%\cite{Berger:2009zg}
\bibitem{Berger:2009zg}
  C.~F.~Berger {\it et al.},
  %``Precise Predictions for W+3 Jet Production at Hadron Colliders,''
  Phys.\ Rev.\ Lett.\  {\bf 102 } (2009)  222001.
  [arXiv:0902.2760 [hep-ph]].
  
%\cite{Berger:2009ep}
\bibitem{Berger:2009ep}
  C.~F.~Berger {\it et al.},
  %``Next-to-Leading Order QCD Predictions for W+3-Jet Distributions at Hadron Colliders,''
  Phys.\ Rev.\  {\bf D80 } (2009)  074036.
  [arXiv:0907.1984 [hep-ph]].
  
%\cite{Ellis:2009zw}
\bibitem{Ellis:2009zw}
  R.~K.~Ellis, K.~Melnikov, G.~Zanderighi,
  %``Generalized unitarity at work: First NLO QCD results for hadronic W+3 jet production,''
  JHEP {\bf 0904 } (2009)  077.
  [arXiv:0901.4101 [hep-ph]].

%\cite{KeithEllis:2009bu}
\bibitem{KeithEllis:2009bu}
  R.~K.~Ellis, K.~Melnikov, G.~Zanderighi,
  %``W+3 jet production at the Tevatron,''
  Phys.\ Rev.\  {\bf D80 } (2009)  094002.
  [arXiv:0906.1445 [hep-ph]].

%\cite{Melnikov:2009wh}
\bibitem{Melnikov:2009wh}
  K.~Melnikov, G.~Zanderighi,
  %``W+3 jet production at the LHC as a signal or background,''
  Phys.\ Rev.\  {\bf D81 } (2010)  074025.
  [arXiv:0910.3671 [hep-ph]].
  
\bibitem{bbbb}
%\cite{Binoth:2009rv}
%\bibitem{Binoth:2009rv}
  T.~Binoth, N.~Greiner, A.~Guffanti, J.~Reuter, J.~-P.~.Guillet, T.~Reiter,
  %``Next-to-leading order QCD corrections to pp --> b anti-b b anti-b + X at the LHC: the quark induced case,''
  Phys.\ Lett.\  {\bf B685 } (2010)  293-296.
  [arXiv:0910.4379 [hep-ph]].

\bibitem{tt2j}
%\cite{Bevilacqua:2010ve}
%\bibitem{Bevilacqua:2010ve}
  G.~Bevilacqua, M.~Czakon, C.~G.~Papadopoulos, M.~Worek,
  %``Dominant QCD Backgrounds in Higgs Boson Analyses at the LHC: A Study of pp -> t anti-t + 2 jets at Next-To-Leading Order,''
  Phys.\ Rev.\ Lett.\  {\bf 104 } (2010)  162002.
  [arXiv:1002.4009 [hep-ph]].

\bibitem{Berger:2010vm}
%\cite{Berger:2010vm}
%\bibitem{Berger:2010vm}
  C.~F.~Berger {\it et al.},
  %``Next-to-Leading Order QCD Predictions for Z,gamma^*+3-Jet Distributions at the Tevatron,''
  Phys.\ Rev.\  {\bf D82 } (2010)  074002.
  [arXiv:1004.1659 [hep-ph]].

\bibitem{Melia:2010bm}
  T.~Melia, K.~Melnikov, R.~Rontsch, G.~Zanderighi,
  %``Next-to-leading order QCD predictions for $W^+W^+jj$ production at the LHC,''
  JHEP {\bf 1012 } (2010)  053.
  [arXiv:1007.5313 [hep-ph]].

\bibitem{Denner:2010jp}
  A.~Denner, S.~Dittmaier, S.~Kallweit {\it et al.},
  %``NLO QCD corrections to WWbb production at hadron colliders,''
    [arXiv:1012.3975 [hep-ph]].

\bibitem{Berger:2010zx}
  C.~F.~Berger {\it et al.},
  % ``Precise Predictions for W + 4 Jet Production at the Large Hadron
  % Collider,''
  [arXiv:1009.2338 [hep-ph]].

\bibitem{Frixione:2002ik}
  S.~Frixione, B.~R.~Webber,
  %``Matching NLO QCD computations and parton shower simulations,''
  JHEP {\bf 0206 } (2002)  029.
  [hep-ph/0204244].

\bibitem{Nason:2004rx}
  P.~Nason,
  %``A New method for combining NLO QCD with shower Monte Carlo algorithms,''
  JHEP {\bf 0411 } (2004)  040.
  [hep-ph/0409146].

\bibitem{Frixione:2007vw}
  S.~Frixione, P.~Nason, C.~Oleari,
  %``Matching NLO QCD computations with Parton Shower simulations: the POWHEG method,''
  JHEP {\bf 0711 } (2007)  070.
  [arXiv:0709.2092 [hep-ph]].

\bibitem{Nason:2009ai}
  P.~Nason, C.~Oleari,
  %``NLO Higgs boson production via vector-boson fusion matched with shower in POWHEG,''
  JHEP {\bf 1002 } (2010)  037.
  [arXiv:0911.5299 [hep-ph]].

\bibitem{Kardos}
A. Kardos, C. Papadopoulos and Z. Tr\'ocs\'anyi,
%``Top quark pair production in association with a
% jet with NLO parton showering
arXiv:1101.267v1 [hep-ph].

%\cite{Ellis:2007qk}
\bibitem{Ellis:2007qk}
  R.~K.~Ellis, G.~Zanderighi,
  %``Scalar one-loop integrals for QCD,''
  JHEP {\bf 0802 } (2008)  002.
  [arXiv:0712.1851 [hep-ph]].

\bibitem{Alioli:2010xd}
  S.~Alioli, P.~Nason, C.~Oleari and E. Re,
  %``A general framework for implementing NLO calculations in shower Monte Carlo programs: the POWHEG BOX,''
  JHEP {\bf 1006 } (2010)  043.
  [arXiv:1002.2581 [hep-ph]].

\bibitem{jager}
%\cite{Jager:2009xx}
%\bibitem{Jager:2009xx}
  B.~Jager, C.~Oleari, D.~Zeppenfeld,
  %``Next-to-leading order QCD corrections to W+ W+ jj and W- W- jj production via weak-boson fusion,''
  Phys.\ Rev.\  {\bf D80 } (2009)  034022.
  [arXiv:0907.0580 [hep-ph]].

%\cite{Alioli:2010qp}
\bibitem{Alioli:2010qp}
  S.~Alioli, P.~Nason, C.~Oleari, E.~Re,
  %``Vector boson plus one jet production in POWHEG,''
  JHEP {\bf 1101 } (2011)  095.
  [arXiv:1009.5594 [hep-ph]].

\bibitem{dps}
%\cite{Kulesza:1999zh}
%\bibitem{Kulesza:1999zh}
  A.~Kulesza, W.~J.~Stirling,
  %``Like sign W boson production at the LHC as a probe of double parton scattering,''
  Phys.\ Lett.\  {\bf B475 } (2000)  168-175.
  [hep-ph/9912232];\\
%\cite{Maina:2009sj}
%\bibitem{Maina:2009sj}
  E.~Maina,
  %``Multiple Parton Interactions in Z+4j, W+- W+- + 0/2j and W+ W- + 2j production at the LHC,''
  JHEP {\bf 0909 } (2009)  081.
  [arXiv:0909.1586 [hep-ph]];\\
%  [arXiv:0909.1586 [hep-ph]].
%\cite{Gaunt:2010pi}
  J.~R.~Gaunt, C.~-H.~Kom, A.~Kulesza, W.~J.~Stirling,
  %``Same-sign W pair production as a probe of double parton scattering at the LHC,''
  Eur.\ Phys.\ J.\  {\bf C69 } (2010)  53-65.
  [arXiv:1003.3953 [hep-ph]].

\bibitem{dreiner}
%\cite{Dreiner:2006sv}
%\bibitem{Dreiner:2006sv}
  H.~K.~Dreiner, S.~Grab, M.~Kramer, M.~K.~Trenkel,
  %``Supersymmetric NLO QCD corrections to resonant slepton production and signals at the Tevatron and the CERN LHC,''
  Phys.\ Rev.\  {\bf D75 } (2007)  035003.
  [hep-ph/0611195].
  
\bibitem{han}
%\cite{Han:2009ya}
%\bibitem{Han:2009ya}
  T.~Han, I.~Lewis, T.~McElmurry,
  %``QCD Corrections to Scalar Diquark Production at Hadron Colliders,''
  JHEP {\bf 1001 } (2010)  123.
  [arXiv:0909.2666 [hep-ph]].

\bibitem{maalampi}
%\cite{Maalampi:2002vx}
%\bibitem{Maalampi:2002vx}
  J.~Maalampi, N.~Romanenko,
  %``Single production of doubly charged Higgs bosons at hadron colliders,''
  Phys.\ Lett.\  {\bf B532 } (2002)  202-208.
  [hep-ph/0201196].

\bibitem{BG}
%\cite{Berends:1987me}
%\bibitem{Berends:1987me}
  F.~A.~Berends, W.~T.~Giele,
  %``Recursive Calculations for Processes with n Gluons,''
  Nucl.\ Phys.\  {\bf B306 } (1988)  759.
  
\bibitem{mint}
  P.~Nason,
  %``MINT: A Computer program for adaptive Monte Carlo integration and generation of unweighted distributions,''
    [arXiv:0709.2085 [hep-ph]].

\bibitem{Cacciari:2008gp}
  M.~Cacciari, G.~P.~Salam, G.~Soyez,
  %``The Anti-k(t) jet clustering algorithm,''
  JHEP {\bf 0804 } (2008)  063.
  [arXiv:0802.1189 [hep-ph]].

\bibitem{Martin:2009iq}
  A.~D.~Martin, W.~J.~Stirling, R.~S.~Thorne, G.~Watt,
  %``Parton distributions for the LHC,''
  Eur.\ Phys.\ J.\  {\bf C63 } (2009)  189-285.
  [arXiv:0901.0002 [hep-ph]].

\bibitem{Sjostrand:2006za}
  T.~Sjostrand, S.~Mrenna, P.~Z.~Skands,
  %``PYTHIA 6.4 Physics and Manual,''
  JHEP {\bf 0605 } (2006)  026.
  [hep-ph/0603175].

\bibitem{Corcella:2000bw}
  G.~Corcella {\it et al.},
  %``HERWIG 6: An Event generator for hadron emission reactions with interfering gluons (including supersymmetric processes),''
  JHEP {\bf 0101 } (2001)  010.
  [hep-ph/0011363].


\end{thebibliography}
\end{document}